\documentstyle[twoside,epsfig]{article}

\catcode`\@=11
\long\def\@makefntext#1{
\protect\noindent \hbox to 3.2pt {\hskip-.9pt  
$^{{\eightrm\@thefnmark}}$\hfil}#1\hfill}		

\def\@makefnmark{\hbox to 0pt{$^{\@thefnmark}$\hss}}	
	
\def\ps@myheadings{\let\@mkboth\@gobbletwo
\def\@oddhead{\hbox{}
\rightmark\hfil\eightrm\thepage}   
\def\@oddfoot{}\def\@evenhead{\eightrm\thepage\hfil
\leftmark\hbox{}}\def\@evenfoot{}
\def\sectionmark##1{}\def\subsectionmark##1{}}

\oddsidemargin=\evensidemargin
\newcounter{sectionc}\newcounter{subsectionc}\newcounter{subsubsectionc}
\renewcommand{\section}[1] {\vspace{12pt}\addtocounter{sectionc}{1} 
\setcounter{subsectionc}{0}\setcounter{subsubsectionc}{0}\noindent 
	{\tenbf\thesectionc. #1}\par\vspace{5pt}}
\renewcommand{\subsection}[1] {\vspace{12pt}\addtocounter{subsectionc}{1} 
	\setcounter{subsubsectionc}{0}\noindent 
	{\bf\thesectionc.\thesubsectionc. {\kern1pt \bfit #1}}\par\vspace{5pt}}
\renewcommand{\subsubsection}[1] {\vspace{12pt}\addtocounter{subsubsectionc}{1}
	\noindent{\tenrm\thesectionc.\thesubsectionc.\thesubsubsectionc.
	{\kern1pt \tenit #1}}\par\vspace{5pt}}
\newcommand{\nonumsection}[1] {\vspace{12pt}\noindent{\tenbf #1}
	\par\vspace{5pt}}

\newcounter{appendixc}
\newcounter{subappendixc}[appendixc]
\newcounter{subsubappendixc}[subappendixc]
\renewcommand{\thesubappendixc}{\Alph{appendixc}.\arabic{subappendixc}}
\renewcommand{\thesubsubappendixc}
	{\Alph{appendixc}.\arabic{subappendixc}.\arabic{subsubappendixc}}

\renewcommand{\appendix}[1] {\vspace{12pt}
        \refstepcounter{appendixc}
        \setcounter{figure}{0}
        \setcounter{table}{0}
        \setcounter{lemma}{0}
        \setcounter{theorem}{0}
        \setcounter{corollary}{0}
        \setcounter{definition}{0}
        \setcounter{equation}{0}
        \renewcommand{\thefigure}{\Alph{appendixc}.\arabic{figure}}
        \renewcommand{\thetable}{\Alph{appendixc}.\arabic{table}}
        \renewcommand{\theappendixc}{\Alph{appendixc}}
        \renewcommand{\thelemma}{\Alph{appendixc}.\arabic{lemma}}
        \renewcommand{\thetheorem}{\Alph{appendixc}.\arabic{theorem}}
        \renewcommand{\thedefinition}{\Alph{appendixc}.\arabic{definition}}
        \renewcommand{\thecorollary}{\Alph{appendixc}.\arabic{corollary}}
        \renewcommand{\theequation}{\Alph{appendixc}.\arabic{equation}}
        \noindent{\tenbf Appendix \theappendixc #1}\par\vspace{5pt}}
\newcommand{\subappendix}[1] {\vspace{12pt}
        \refstepcounter{subappendixc}
        \noindent{\bf Appendix \thesubappendixc. {\kern1pt \bfit #1}}
	\par\vspace{5pt}}
\newcommand{\subsubappendix}[1] {\vspace{12pt}
        \refstepcounter{subsubappendixc}
        \noindent{\rm Appendix \thesubsubappendixc. {\kern1pt \tenit #1}}
	\par\vspace{5pt}}

\newcommand{\textlineskip}{\baselineskip=13pt}
\newcommand{\smalllineskip}{\baselineskip=10pt}

\def\eightcirc{
\begin{picture}(0,0)
\put(4.4,1.8){\circle{6.5}}
\end{picture}}
\def\eightcopyright{\eightcirc\kern2.7pt\hbox{\eightrm c}}

\def\abstracts#1#2#3{{
	\centering{\begin{minipage}{4.5in}\baselineskip=10pt\footnotesize
	\parindent=0pt #1\par 
	\parindent=15pt #2\par
	\parindent=15pt #3
	\end{minipage}}\par}} 

\newcommand{\bibit}{\nineit}

\renewenvironment{thebibliography}[1]
	{\frenchspacing
	 \ninerm\baselineskip=11pt
	 \begin{list}{\arabic{enumi}.}
	{\usecounter{enumi}\setlength{\parsep}{0pt}
	 \setlength{\leftmargin 12.7pt}{\rightmargin 0pt} 
	 \setlength{\itemsep}{0pt} \settowidth
	{\labelwidth}{#1.}\sloppy}}{\end{list}}

\newcounter{itemlistc}
\newcounter{romanlistc}
\newcounter{alphlistc}
\newcounter{arabiclistc}

\newcommand{\fcaption}[1]{
        \refstepcounter{figure}
        \setbox\@tempboxa = \hbox{\footnotesize Fig.~\thefigure. #1}
        \ifdim \wd\@tempboxa > 5in
           {\begin{center}
        \parbox{5in}{\footnotesize\smalllineskip Fig.~\thefigure. #1}
            \end{center}}
        \else
             {\begin{center}
             {\footnotesize Fig.~\thefigure. #1}
              \end{center}}
        \fi}

\newcommand{\tcaption}[1]{
        \refstepcounter{table}
        \setbox\@tempboxa = \hbox{\footnotesize Table~\thetable. #1}
        \ifdim \wd\@tempboxa > 5in
           {\begin{center}
        \parbox{5in}{\footnotesize\smalllineskip Table~\thetable. #1}
            \end{center}}
        \else
             {\begin{center}
             {\footnotesize Table~\thetable. #1}
              \end{center}}
        \fi}

\def\@citex[#1]#2{\if@filesw\immediate\write\@auxout
	{\string\citation{#2}}\fi
\def\@citea{}\@cite{\@for\@citeb:=#2\do
	{\@citea\def\@citea{,}\@ifundefined
	{b@\@citeb}{{\bf ?}\@warning
	{Citation `\@citeb' on page \thepage \space undefined}}
	{\csname b@\@citeb\endcsname}}}{#1}}

\newif\if@cghi
\def\cite{\@cghitrue\@ifnextchar [{\@tempswatrue
	\@citex}{\@tempswafalse\@citex[]}}
\def\citelow{\@cghifalse\@ifnextchar [{\@tempswatrue
	\@citex}{\@tempswafalse\@citex[]}}
\def\@cite#1#2{{$\null^{#1}$\if@tempswa\typeout
	{IJCGA warning: optional citation argument 
	ignored: `#2'} \fi}}

\def\pmb#1{\setbox0=\hbox{#1}
	\kern-.025em\copy0\kern-\wd0
	\kern.05em\copy0\kern-\wd0
	\kern-.025em\raise.0433em\box0}

\def\fnt#1#2{\footnotetext{\kern-.3em
	{$^{\mbox{\scriptsize #1}}$}{#2}}}

\def\fpage#1{\begingroup
\voffset=.3in
\thispagestyle{empty}\begin{table}[b]\centerline{\footnotesize #1}
	\end{table}\endgroup}

\def\runninghead#1#2{\pagestyle{myheadings}
\markboth{{\protect\footnotesize\it{\quad #1}}\hfill}
{\hfill{\protect\footnotesize\it{#2\quad}}}}
\font\tenrm=cmr10
\font\tenit=cmti10 
\font\tenbf=cmbx10
\font\bfit=cmbxti10 at 10pt
\font\ninerm=cmr9
\font\nineit=cmti9

\font\eightrm=cmr8

\def\qed{\hbox{${\vcenter{\vbox{			
   \hrule height 0.4pt\hbox{\vrule width 0.4pt height 6pt
   \kern5pt\vrule width 0.4pt}\hrule height 0.4pt}}}$}}

\begin{document}

\runninghead{The CLEO III RICH Detector}{The CLEO III RICH Detector}

\normalsize\textlineskip
\thispagestyle{empty}
\setcounter{page}{1}


\vspace*{0.88truein}

\fpage{1}
\centerline{\bf The CLEO III RICH Detector}
\vspace*{0.37truein}
\centerline{\footnotesize{GOBINDA MAJUMDER}}
\vspace*{0.015truein}
\centerline{\footnotesize\it Physics Department, Syracuse University}
\baselineskip=10pt
\centerline{\footnotesize\it Syracuse, N.~Y.~13244-1130, USA}
\vspace*{0.225truein}

\vspace*{0.21truein}
\abstracts{CLEO III upgrade was completed with the integration of 
Ring Imaging CHerenkov(RICH) detector for charged particle 
identification. The design of this cylindrical detector consists of LiF
crystal radiators and multi-wire proportional chamber photon detectors 
coupled through a $N_2$ filled expansion gap. Early performance
on $K/\pi$ separation is presented.}{}{} 


\vspace*{1pt}\textlineskip
\section{Introduction}
\vspace*{-0.5pt}
\noindent
 In the past 20 years of its existence, the CLEO detector has been
associated with many discoveries and precision measurements of heavy
flavour physics.
The latest generation of the detector, the CLEO III, has been upgraded.
One of the new
features of the CLEO III detector is it's particle identification
system, the Ring Imaging CHerenkov (RICH) detector.

  The basic principle of the differential Cherenkov detector is, given by 
the  relation $\Delta\,cos^2\theta\,=\,\Delta\,m^2/(n^2\,p^2)$,
where $n$ is the 
refractive index of LiF, $p$ is momentum and $\Delta\,m^2$ is the 
mass square difference between two particles. 
In order to achieve
efficient particle identification with low fake rates, a  design goal
of 4$\sigma$ K/$\pi$ separation at 2.65\,GeV was set. 
This momentum is given by two-body B decays.
At this momentum,
difference between K/$\pi$ Cherenkov angle, $\Delta\theta_{\gamma}$ = 
14.4 mrad. Along with the 
1.8$\sigma$ $dE/dx$ identification from the drift chamber, one requires
a per track angular resolution of $\sigma_{\theta_{trk}}$ = 4\,mrad. Using a naive
approximation 
$\sigma_{\theta_{trk}}\,=\,\sigma_{\theta_{\gamma}}/\sqrt{N_{\gamma}}$, 
the benchmark of RICH design was therefore to have an average of 12 
photons with a resolution of 14\,mrad per photon.

\section{Detector design}
  CLEO RICH detector design was mainly constrained by two factors, (a)
only 20\,cm of radial space available between the drift 
chamber and the calorimeter with 2.5\,m of length (80\% of solid angle) and
(b) for good performance of the EM calorimeter 
RICH material had to be less
than 15\% of the radiation length.

  The overall RICH design is a cylinder with compact photon detector 
modules on the outer layer and radiator crystals on the inner layer,
forming a
12$^{\circ}$ sector in azimuth.
The driving constraints for the 
design in many respects was the choice of Triethylamine (TEA) as a 
photon-absorber, as it is chemically aggressive and also shows a high
quantum efficiency in the VUV regime (135$-$165\.nm). 

{\bf Crystal Radiators:}
  The basic design consists of 30 rows of crystal along the axis of 
cylinder with each row containing 14 LiF crystals, of dimension 
$\rm \sim\:175 \times 172 \times 10 \:mm^3$ mounted on a carbon fiber 
cylinder. Due to high refractive index of LiF (n $\sim$ 1.5), a large 
number of cherenkov photons are internally reflected, resulting in partial 
images. At the normal track incidence 
all photons are reflected back due to total internal
reflection. In order to fix this situation, a novel radiator geometry
is used in the central region (120 crystal). The light emitting 
surface of the radiator is cut with a profile resembling the teeth of 
a saw, and therefore is referred to as the sawtooth radiator.\cite{1}

{\bf Expansion volume:}
  This is essentially an empty space, 157mm in radial distance, filled 
up with pure N$_2$ gas. O$_2$ and H$_2$O contamination is kept less
than 10\,ppm to avoid any loss of photons in this volume. Transmission
efficiency is more 99\% at 150\,nm.

{\bf Photon detectors:}
  These are photosensitive asymmetric multiwire proportional chambers,
filled with CH$_4$ carrier gas bubbled
through liquid TEA at 15$^\circ$ (5.5\% vapour concentration) and 
having 2mm thick CaF$_2$ windows with 100$\mu m$ thin Ag strip
providing one of the cathode planes. TEA has 
a peak QE of $\sim$ 33\% at 150\,nm and a bandwidth 
of 135-165\,nm.\cite{2} Anode wires are made of 20$\mu m$ diameter Au-W and 
the array of 8.0$\times$7.53\,mm$^2$ cathode pads on outer surface
is used to collect induced image charge.
This is an optimal design to maximise 
photon conversion efficiency ($\ell_{abs}$ = 0.56\,mm at 150\,nm) and 
anode-cathode charge coupling by having the wire-pad gap be as small 
as 1\,mm. 

{\bf Readout Electronics:}
  The choice of readout electronics is governed by the exponential 
charge distribution of single photoelectron avalanche and also by the 
time allowed for the readout. Analog front-end electronics with low 
noise and high dynamic range are used, which has very low rms noise
ENC = 130e$^-$ + 9$C_{det}e^-$/pF $\approx~150e^-$
for modest $C_{det} \approx$ 2pF.

The charge information is necessary to determine the centroid of the 
photoelectron as well as disentangle the overlap of two nearby photons.
The latter requires high segmentation. Given the modest pass size over
large detector area, there are 230,400 pads in total.
Sparsification works well 
since occupancy is low ($<$1\%).

\section{Performance}
  The RICH detector was installed along with other subdetectors in CLEO in 
Aug 1999 and the first data were collected starting from Nov 1999. The 
detector performance has been satisfactory. The transparency of expansion 
volume is more than 99\% at 150\,nm. The stable high voltage system 
has been successfully operated at different voltage setting to choose 
optimal operation voltage. The average detected 
charge per Cherenkov photon is
$\sim 2.2 \times 10^4$ electrons. 
The electronic noise has a mean value
$\sim 500\,e^-$, with common mode subtraction and $\sim 790\,e^-$ 
including the coherent component.

  During the engineering run, preliminary study of detector performance
was mainly based on the Bhabha events. The angular resolution is very close 
to what was expected. Some observed deterioration 
is due to noise and lack of 
proper alignment. The results are shown in Table ~\ref{tabresult}.

\begin{table}
\begin{center}
\begin{tabular}{|ccccccc|}  \hline
  & \multicolumn{2}{c}{$\sigma_{\theta_{\gamma}}$}  
  & \multicolumn{2}{c}{N$_{\gamma}$} 
  & \multicolumn{2}{c|}{$\sigma_{\theta_{trk}}$} \\ \hline
  & Flat &Sawtooth & Flat &Sawtooth & Flat &Sawtooth \\
MC   & 14.0 & 12.2 & 10.6 & 14.1 & 4.96 & 3.39 \\
Data & 14.0 & 12.9 & 10.2 & 12.0 & 5.27 & 4.21 \\
\hline \end{tabular}
\end{center}
\tcaption{\label{tabresult} Comparison of RICH Data with MC expectation.}
\end{table}

\begin{figure}[htbp]
\epsfig{figure=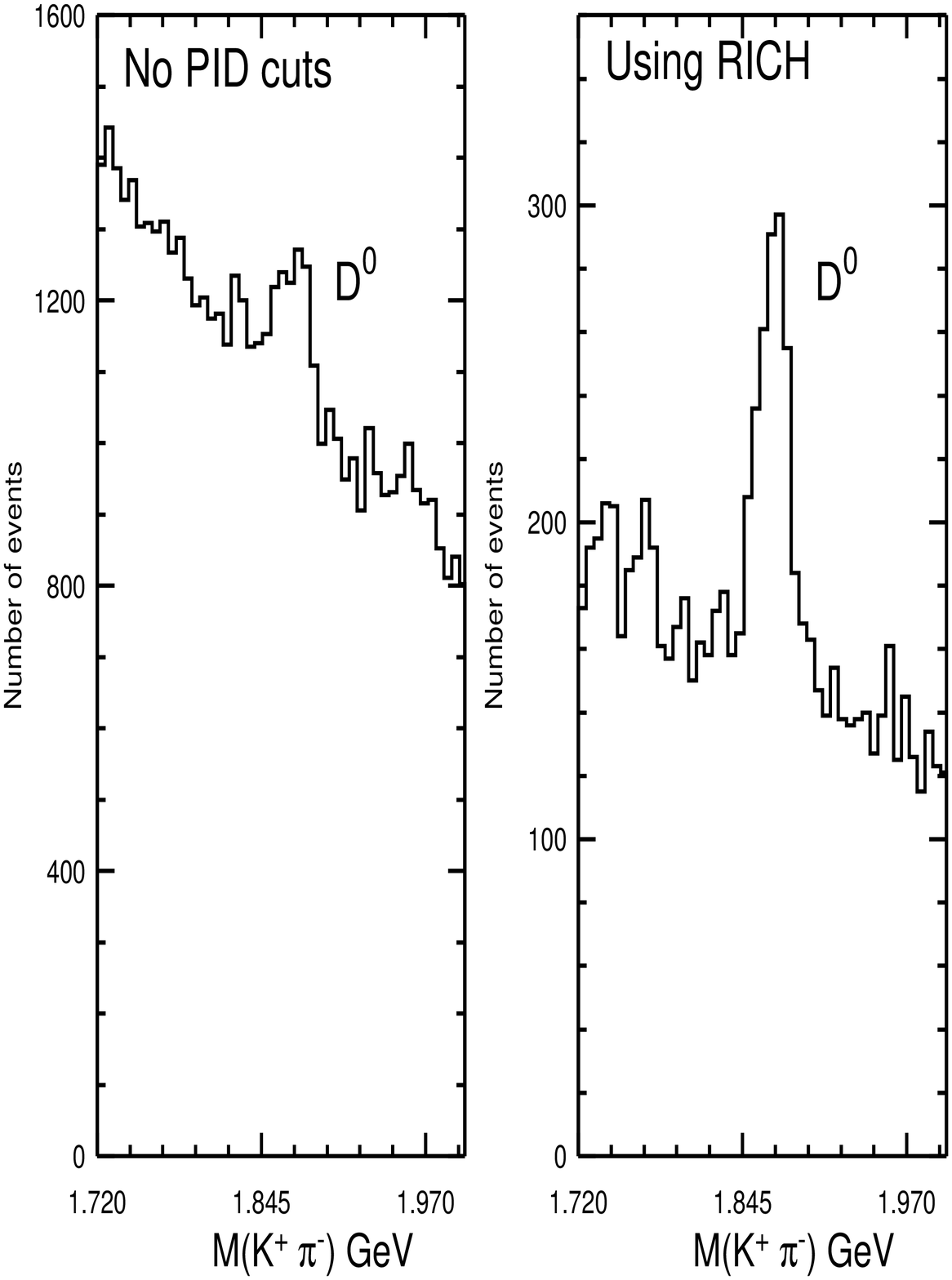, height=1.5in, width=2.45in}
\epsfig{figure=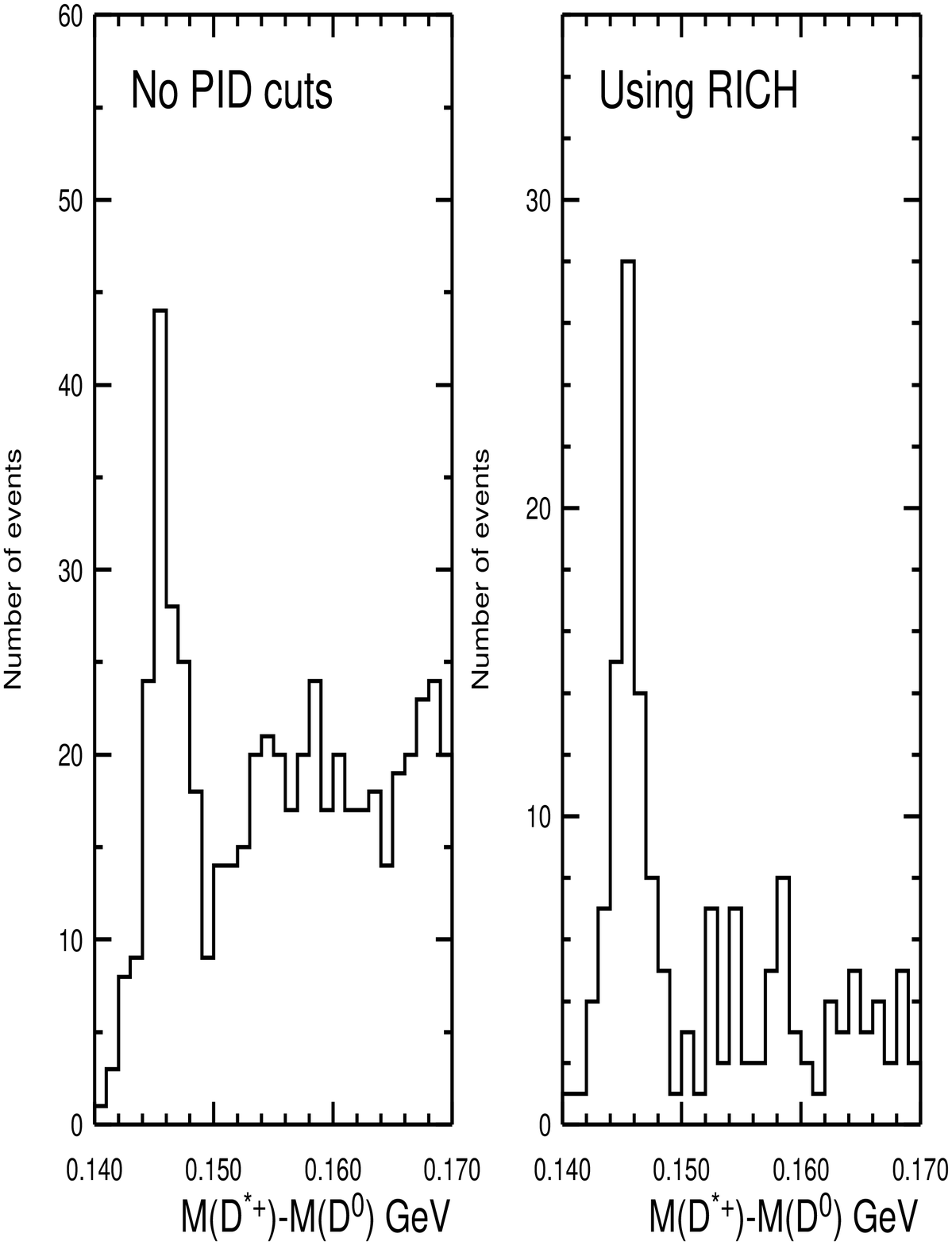, height=1.5in, width=2.45in}
\fcaption{\label{d0plot} Efficiency of RICH detector to reconstruct
$D^{\star\pm}$ and $D^{\circ}$.}
\end{figure}

 As a first attempt at physics analysis,
$D^\circ \rightarrow K^-\pi^+$  and 
$D^{\star\pm} \rightarrow K^-\pi^+\pi^\pm$ were reconstructed. For 80\% 
efficiency of $D^\circ$ in the signal region, RICH could reduce the 
background by an order of magnitude. Similarly $D^{\star\pm}$ was 
reconstructed, removing virtually all background from 
M$_D^{\star\pm}-$M$_D^{\circ}$ peak. These distributions are shown in
figure ~\ref{d0plot}.

 These results are preliminary. Using better tracking 
the Cherenkov angle resolution should improve.

\section{Conclusion}
  The engineering run has proven the satisfactory performance of the RICH
detector, which fulfilled its benchmark. Since Aug 2000, CLEO has been 
taking physics data. The RICH detector will be an important
element for the upcoming CLEO physics results.

\nonumsection{References}

\end{document}